\documentclass[onecollarge,natbib]{svjour3}
\bibpunct{[}{]}{;}{n}{}{,} % to get "[numbered]" references from natbib
\smartqed  % flush right qed marks, e.g. at end of proof
\usepackage{graphicx}
\begin{document}

\title{\bf The MAID Legacy and Future}

\author{Lothar Tiator}

\institute{L. Tiator \at
              Universit{\"a}t Mainz \\
              \email{tiator@uni-mainz.de}
}

\date{Received: date / Accepted: date}

\maketitle

\begin{abstract}
The MAID project is a collection of theoretical models for
pseudoscalar meson photo- and electroproduction from nucleons. It is
online available and produces results in real time calculations. In
addition to kinematical variables also model parameters, especially
for baryon resonances, can be online changed and investigated. Over
20 years MAID has become quite popular and the MAID web pages have
been called more than 7.7 million times. 
\end{abstract}

\section{Introduction}
\label{intro} After the exploration of the baryonic spectrum with
pion scattering in the 1980s, photo- and electroproduction of
mesons, mainly at Mainz, Bonn and JLab have become the main source
of further investigations of $N$ and $\Delta$ resonances. Among
them, pion electroproduction is the main source for investigations
of the transition form factors of the nucleon to excited $N$ and
$\Delta$ baryons. In addition also two-pion and eta
electroproduction have been very useful for studies of selected
nucleon resonances. After early measurements of the $G_M^*$ form
factor of the $N\rightarrow\Delta(1232)$ transition already in the
1960s, in the 1990s a large program was running at Mainz, Bonn,
Bates and JLab in order to measure the very small $E/M$ ratio of the
$N\rightarrow\Delta$ transition and the $Q^2$ dependence of the
$E/M$ and $S/M$ ratios in order to get information on the internal
quadrupole deformation of the nucleon and the $\Delta$. In parallel
large progress was achieved in various kinds of quark models that
gave predictions to $N\rightarrow N^*$ and $N\rightarrow \Delta^*$
transition form factors. Only at JLab both the energy and the photon
virtuality were available to measure those form factors for a set of
nucleon resonances up to $Q^2\approx 5$~GeV$^2$. Two recent review
articles on the electromagnetic excitation of nucleon resonances
give a very good overview over experiment and theory and latest
developments~\cite{Tiator:2011pw,Aznauryan:2011qj}.

\section{The MAID project}

The first MAID program appeared in 1998 for pion photo- and
electroproduction on the nucleon~\cite{Maid98}. It was extended and
updated with new data in the years 2000 and 2003. A major update was
published in 2007~\cite{MAID07} and covers an energy range from
threshold until $W=2$~GeV and photon virtualities up to
$Q^2=5$~GeV$^2$. A whole series of transition form factors was
analyzed both as single-$Q^2$ data points and in a $Q^2$-dependent
analysis with simple polynomial and exponential
parameterizations~\cite{Tiator:2011pw}. Soon afterwards the
Dubna-Mainz-Taipei (DMT) dynamical model~\cite{Kamalov:2000en} was
going online. The DMT model has since proven an enormous predictive
power for pion photo- and electroproduction from threshold up to the
$\Delta(1232)$ resonance. Beyond the $\Delta(1232)$ the DMT model is
very similar to the unitary isobar model MAID. At the same time also
the isobar models KaonMAID~\cite{Mart:1999ed} and
EtaMAID~\cite{Chiang:2001as} were developed. The online KaonMAID has
not yet been updated, but extensions and updates have been published
over the time until very recently~\cite{Clymton:2017nvp}. The
EtaMAID was extended in 2003 with a Regge background
approach~\cite{Chiang:2002vq}. Also EtaprimeMAID2003 was
established~\cite{Chiang:2002vq}, but was not further updated until
now. In 2007 the TwoPionMAID joined the project, an isobar model for
two-pion photoproduction on the nucleon~\cite{Fix:2005if}. Finally,
a chiral effective theory approach, ChiralMAID~\cite{Hilt:2013fda},
where all low-energy constants were fitted to the world data of pion
threshold production, was added, which is applicable for neutral and
charged pion photo- and electroproduction in the threshold region
until the onset of the $\Delta(1232)$ resonance and photon
virtualities of $Q^2\le 0.15$~GeV$^2$.
\begin{figure}[h]
\begin{center}
\includegraphics[width=12.0cm]{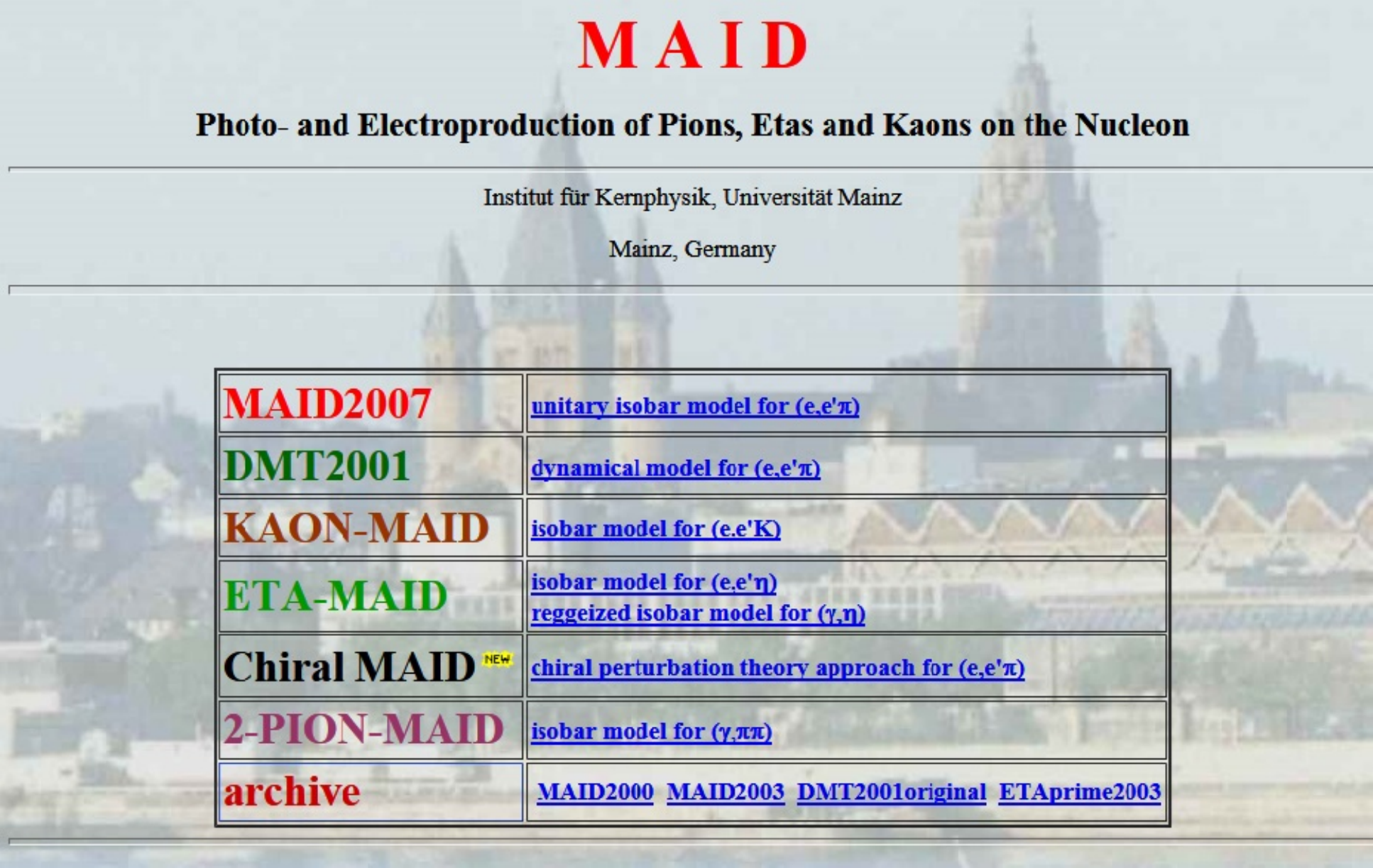}
\vspace{3mm} \caption{\label{fig:maid} The MAID project on the Mainz
web site {\texttt https://maid.kph.uni-mainz.de/}. Since the start
of MAID98 in 1998, the programs have been called more than 7.7
million times.}
\end{center}
\end{figure}
The welcome page of the MAID programs on {\texttt
https://maid.kph.uni-mainz.de/} is shown in Fig.~\ref{fig:maid}.

\begin{figure}[h]
\begin{center}
\includegraphics[width=12.0cm]{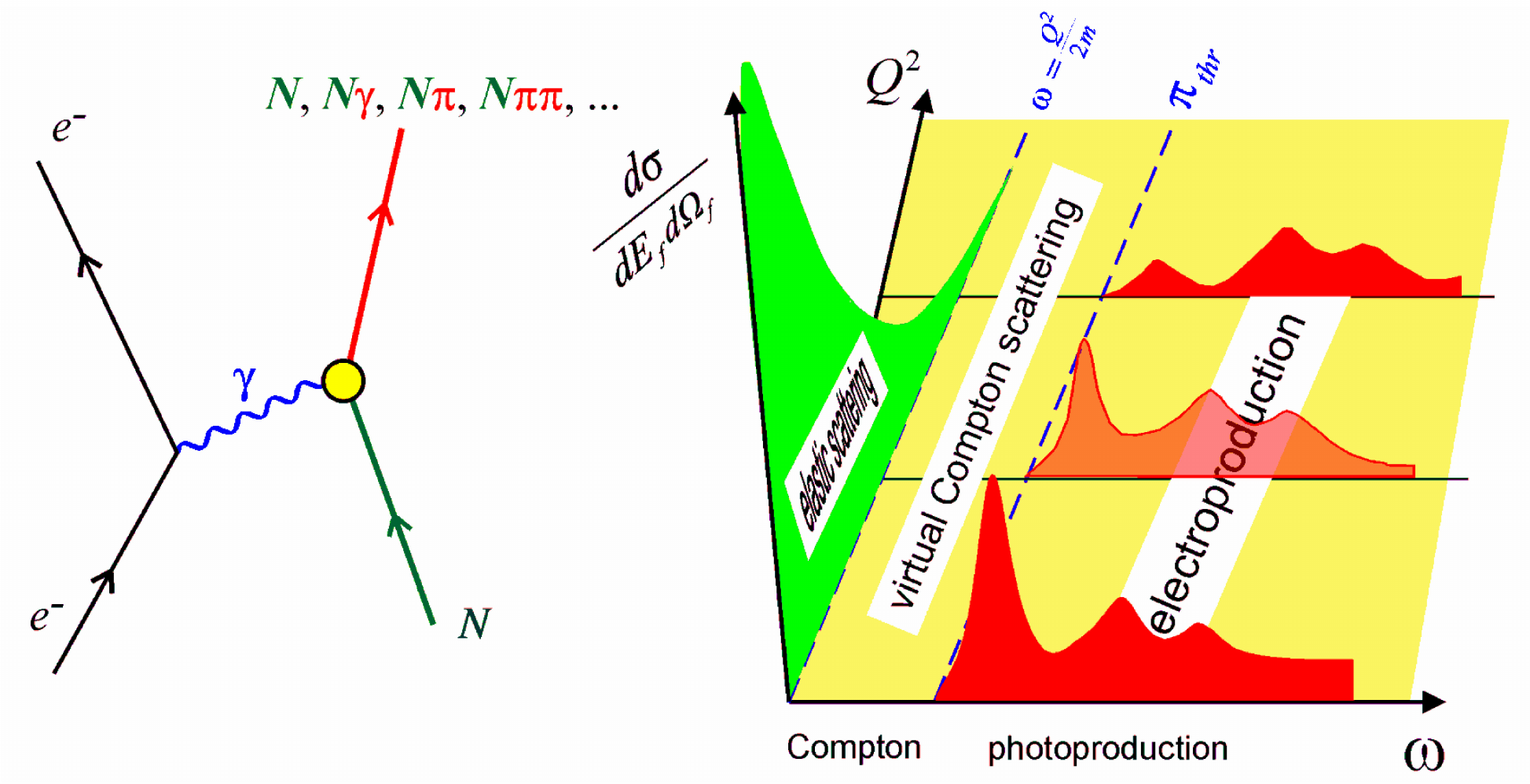}
\vspace{3mm} \caption{\label{fig:response} Meson electroproduction
and space-like baryon resonance excitations.}
\end{center}
\end{figure}

In Fig.~\ref{fig:response} a sketch of the nucleon response to
space-like virtual photons is shown. Besides elastic scattering and
real- and virtual Compton scattering, the meson production processes
dominate the response. At low $Q^2$ the $\Delta(1232)$ is the major
player, but its contribution drops faster than the dipole form
factor, therefore other resonances are more pronounced at higher
$Q^2$. In the time-like region meson production and transition form
factors are also accessible by experiment via the Dalitz decays
until the so-called pseudo-threshold, $Q^2_{pt}=-(M_{N^*}-M_N)^2$,
which is at $\{-0.086, -0.252, -0.55\}$~GeV$^2$ at the resonance
positions of the $\Delta(1232)3/2^+, N(1440)1/2^+, N(1680)5/2^+$
resonances, respectively. Beyond the pseudo-threshold, the time-like
region becomes unphysical and opens again at
$Q^2_{pt}=-(M_{N^*}+M_N)^2< -4$~GeV$^2$.

\section{The MAID ansatz}

In the spirit of a dynamical approach to pion photo- and
electroproduction, the $t$-matrix of the unitary isobar model MAID
is set up by the ansatz~\cite{Tiator:2011pw,MAID07}

\begin{equation}
t_{\gamma\pi}(W)=t_{\gamma\pi}^B(W) + t_{\gamma\pi}^{R}(W)\,,\label{eq:DM}
\end{equation}
with a background and a resonance $t$-matrix, each of them
constructed in a unitary way. Of course, this ansatz is not unique.
However, it is a very important prerequisite to clearly separate
resonance and background amplitudes within a Breit-Wigner concept
also for higher and overlapping resonances.

For a specific partial wave $\alpha = \{j,l,\ldots\}$, the
background $t$-matrix is set up by a potential multiplied by the
pion-nucleon scattering amplitude in accordance with the K-matrix
approximation,
\begin{equation}
 t^{B,\alpha}_{\gamma\pi}(W,Q^2)=v^{B,\alpha}_{\gamma\pi}(W,Q^2)\,[1+it_{\pi N}^{\alpha}(W) ]\, ,
\label{eq:Kmatrix}
\end{equation}
where only the on-shell part of pion-nucleon rescattering is
maintained and the off-shell part from pion-loop contributions is
neglected. Whereas this approximation would fail near the threshold
for $\gamma,\pi^0$, it is well justified in the resonance region
because the main contribution from pion-loop effects is absorbed by
the nucleon resonance dressing.

The background potential $v_{\gamma\pi}^{B,\alpha}(W,Q^2)$ is
described by Born terms obtained with an energy-dependent mixing of
pseudovector-pseudoscalar $\pi NN$ coupling and $t$-chan\-nel vector
meson exchanges. The mixing para\-meters and coupling constants are
determined by an analysis of non-resonant multipoles in the
appropriate energy regions~\cite{Maid98}.
In the latest version MAID2007~\cite{MAID07}, the $S$, $P$, $D$, and
$F$ waves of the background contributions are unitarized as
explained above, with the pion-nucleon elastic scattering
amplitudes, $t^{\alpha}_{\pi N}=[\eta_{\alpha}
\exp(2i\delta_{\alpha})-1]/2i$, described by phase shifts
$\delta_{\alpha}$ and the inelasticity parameters $\eta_{\alpha}$
taken from the GWU/SAID analysis~\cite{Arndt:1995ak}.

For the resonance contributions Breit-Wigner forms for the
resonance shape are assumed, following Ref.~\cite{Maid98}
\begin{eqnarray}
t_{\gamma\pi}^{R,\alpha}(W,Q^2) &=& {\bar{\cal A}}_{\alpha}^R(W,Q^2)\,
 \frac{f_{\gamma N}(W)\Gamma_{tot}(W)\,M_R\,f_{\pi N}(W)}{M_R^2-W^2-iM_R\,
\Gamma_{tot}(W)}\, e^{i\phi_R(W)}\,, \label{eq:BW}
\end{eqnarray}
where $f_{\pi N}(W)$ is the usual Breit-Wigner factor describing the
decay of a resonance with total width $\Gamma_{tot}(W)$, partial
$\pi N$ width $\Gamma_{\pi N}(W)$, and spin $j$,
\begin{equation}
f_{\pi N}(W)=C_{\pi N}\left[\frac{1}{(2j+1)\pi}\frac{\kappa(W)}{q(W)}
\frac{M_N}{M_R}\frac{\Gamma_{\pi N}(W)}{\Gamma_{\rm {tot}}^2(W)}\right]^{1/2}\,. \label{eq:fpin}
\end{equation}

The energy dependence of the partial widths and of the $\gamma NN^*$
vertex can be found in Ref.~\cite{MAID07}.
The unitary phase $\phi_R(W)$ in Eq.~(\ref{eq:BW}) is introduced to
adjust the total phase such that the Fermi-Watson theorem is
fulfilled below two-pion threshold. In the inelastic region above
the two-pion threshold it can be considered as a free parameter.

While the original version of MAID included only the 7 most
important nucleon resonances with only transverse e.m. couplings in
most cases, MAID2007 describes all 13 four-star resonances below
$W=2$~GeV. In a forthcoming update of MAID, these list of dominant
resonances is no more sufficient, due to the high accuracy of the
data and the availability of many polarization observables with
single and double polarization. With these data also the weaker
three-star and two-star resonances can be analyzed and will be
included in the model.

\section{Transition form factors}

In most cases, the resonance couplings $\bar{\mathcal
A}_{\alpha}^R(W,Q^2)$ are assumed to be independent of the total
energy. However, an energy dependence may occur if the resonance is
parameterized in terms of the virtual photon three-momentum
$k(W,Q^2)$, e.g., in MAID2007 for the $\Delta(1232)$ resonance. For
all other resonances a simple $Q^2$ dependence is assumed for
$\bar{\mathcal A}_{\alpha}(Q^2)$. These resonance couplings are
taken as constants for a single-Q$^2$ analysis, e.g., for
photoproduction ($Q^2=0$) but also at any fixed $Q^2>0$, whenever
sufficient data with W and $\theta$ variation are available.
Independently from this single-Q$^2$ analysis, also a
$Q^2$-dependent analysis, with a simple ansatz using polynomials and
exponentials, was performed. In MAID2007 the $Q^2$ dependence of the
e.m. $N\rightarrow\Delta(1232)$ transition form factors is
parameterized as follows:
\begin{eqnarray}
G_{E,M}^*(Q^2)&=&g_{E,M}^0 (1 + \beta_{E,M}
Q^{2})  e^{-\gamma_{E,M} Q^2}G_D(Q^2)\,\label{eq:GEMstar} \,,\\
G_{C}^*(Q^2)&=&g_{C}^0 \frac{1 + \beta_{C}Q^{2}}{1+\delta_C
Q^2/(4M_N^2)} \frac{2 M_\Delta}{\kappa_\Delta} e^{-\gamma_{C}
Q^2}G_D(Q^2)\,,\label{eq:GCstar}
\end{eqnarray}
where $G_D(Q^2)=1/(1+Q^2/0.71\,{\rm {GeV}}^2)^2$ is the dipole form
factor and the parameters are given in Table~\ref{tab:GMEC}.

\begin{table}[htbp]
\caption{Parameters for the $N\rightarrow \Delta(1232)$ transition
form factors $G_M^*,G_E^*,G_C^*$ defined by
Eqs.~(\ref{eq:GEMstar}-\ref{eq:GCstar}). The normalization values
$g_\alpha^0$ at the photon point $(Q^2=0)$ and $\delta_\alpha$ are
dimensionless, the parameters $\beta$ and $\gamma$ are given in
GeV$^{-2}$. } \label{tab:GMEC}
\begin{center}
\begin{tabular}{|c|cccc|}
\hline
  & $g_\alpha^0$ & $\beta_\alpha$ & $\gamma_\alpha$  & $\delta_\alpha$\\
\hline
 M1 & 3.00    & 0.0095  &  0.23  &\\
 E2 & 0.0637  & -0.0206 &  0.16  &\\
 C2 &  0.1240 &  0.120  &  0.23  & 4.9\\
\hline
\end{tabular}
\end{center}
\end{table}

For all other $N$ and $\Delta$ resonances the couplings are
parameterized as functions of $Q^2$ by the ansatz
\begin{equation}
\bar{\mathcal A}_{\alpha}(Q^2) =\bar{\mathcal A}_{\alpha}(0) (1+a_1
Q^2+a_2 Q^4 +a_4 Q^8)\, e^{-b_1 Q^2}\,. \label{eq:ffpar}
\end{equation}
For such an ansatz the parameters $\bar{\mathcal A}_{\alpha}(0)$ are
determined in a fit to the world database of photoproduction, and
the parameters $a_i$ and $b_1$ are obtained from a combined fit of
all the electroproduction data at different $Q^2$. The latter
procedure is called the $Q^2$-dependent fit. In MAID the photon
couplings $\bar{\mathcal A}_{\alpha}(0)$ are input parameters,
directly related to the helicity couplings $A_{1/2},\, A_{3/2}$, and
$S_{1/2}$ of nucleon resonance excitation. Relations between these
helicity form factors, Sachs form factors $G_M^*,G_E^*,G_C^*$ and
Dirac form factors $F_1,F_2,F_3$ can be found in
Refs.~\cite{MAID07,Tiator:2011pw}.

\begin{figure}
\begin{center}
\includegraphics[width=12cm]{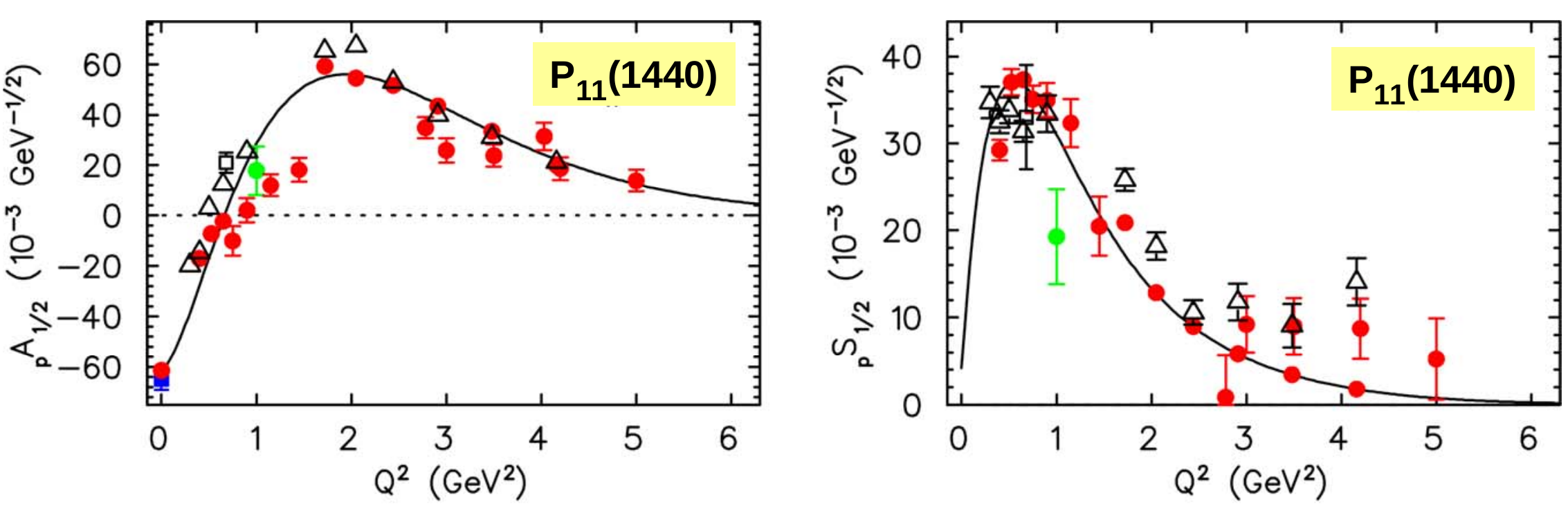}
\vspace*{3mm}
\includegraphics[width=12cm]{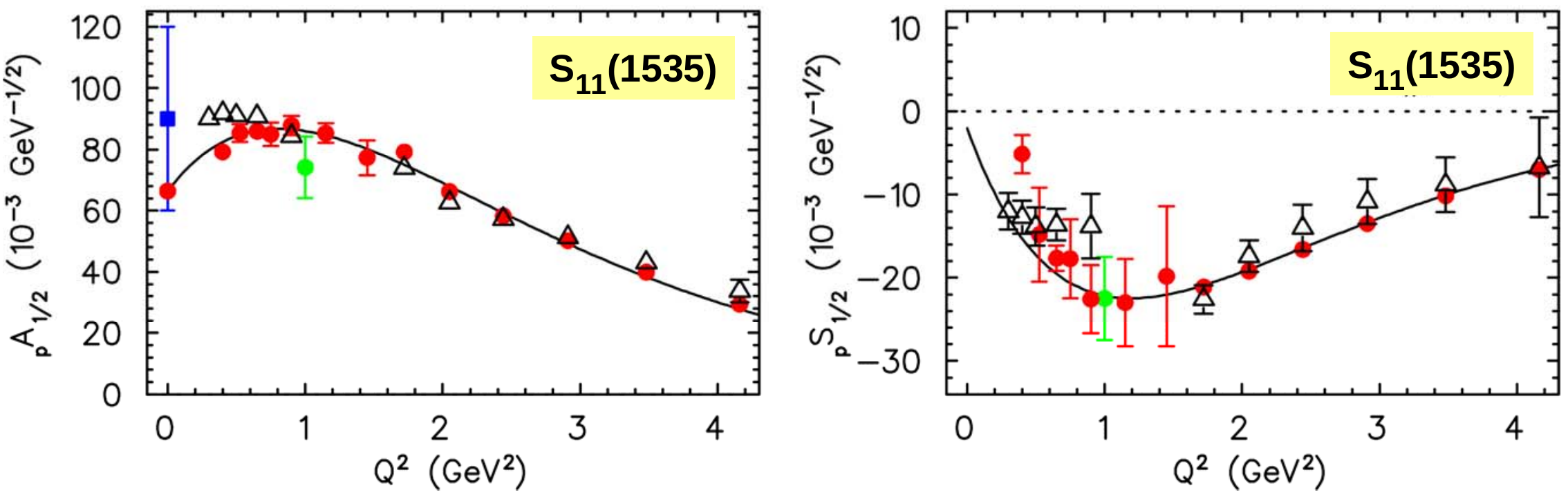}
\vspace{3mm} \caption{\label{fig:s11_p11_ff} Transverse $_pA_{1/2}$
and longitudinal $_pS_{1/2}$ transition form factors of the Roper
$N(1440)1/2^+$ and S11 $N(1535)1/2^-$ nucleon resonances. The red
circles are single-$Q^2$ results from our MAID analyses and the
black triangles from the JLab analysis~\cite{Aznauryan:2009mx}. For
further details see Ref.~\cite{Tiator:2011pw}. }
\end{center}
\end{figure}

In Fig.~\ref{fig:s11_p11_ff} the transverse and longitudinal
helicity form factors for the transitions to the spin $1/2$ nucleon
resonances $N(1440)1/2^+$ (P11, Roper) and $N(1535)1/2^-$ (S11) are
shown. A comparison of the MAID parametrization according to
Eq.~(\ref{eq:ffpar}) with the single-$Q^2$ analyses from
MAID~\cite{MAID07} and CLAS@JLab~\cite{Aznauryan:2009mx} shows a
very good agreement. These transition form factors are well
understood up to $Q^2\approx 5$~GeV$^2$. Whereas the transverse form
factors are constrained by photoproduction at $Q^2=0$, no such
constraint is possible for longitudinal form factors. Therefore, in
the right panels of Fig.~\ref{fig:s11_p11_ff} it remains unclear,
whether the MAID extrapolation is realistic at the photon point.

\begin{figure}
\begin{center}
\includegraphics[width=7cm]{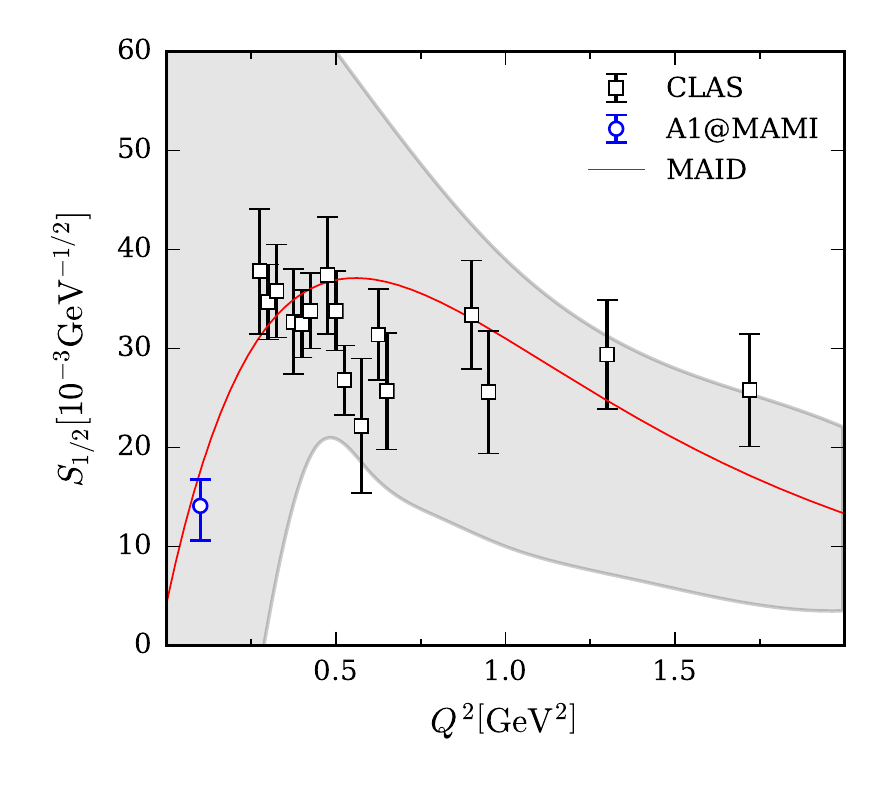}
\vspace{3mm} \caption{\label{fig:roper_s12} Longitudinal transition
form factor $_pS_{1/2}$ of the Roper $N(1440)1/2^+$ resonance. The
red line shows the MAID2007 prediction constrained by the Siegert
Theorem. The black squares are CLAS data~\cite{Aznauryan:2009mx} and
the lowest data point (blue circle) is the new result of a
beam-recoil polarization measurement of A1@MAMI in
Mainz~\cite{Stajner:2017fmh} The gray shaded band shows the
convolution of various model predictions, see
Ref.~\cite{Stajner:2017fmh}. }
\end{center}
\end{figure}

In 2017 a new measurement of the A1@MAMI collaboration in Mainz was
published~\cite{Stajner:2017fmh}, where the longitudinal transition
form factor of the Roper resonance was measured at the lowest
momentum transfer, $Q^2=0.1$~GeV$^2$, see Fig.~\ref{fig:roper_s12}.

A shaded area shows the immense range of predictions for this
longitudinal transition form factor, and in
Refs.~\cite{Aznauryan:2011qj,Stajner:2017fmh} various quark and
meson-baryon models are mentioned, that predict quite different
results. Most of them practically diverge in the $Q^2\rightarrow 0$
region. However, the good agreement between MAID and the new data is
not an accident, but it is due to a constraint from the Siegert
Theorem in the MAID parametrization~\cite{MAID07}. Principally, the
Siegert theorem relates the longitudinal and electric form factors
at pseudo-threshold, which is located in the time-like region at
$Q^2_{pt}=-(M_{N^*}-M_N)^2$, and has a value of $-0.25$~GeV$^2$ for
the Roper transition. The Roper transition, however, does not have
an electric form factor, but as a minimal constraint, all
longitudinal helicity form factors $S_{1/2}(Q^2)$ must vanish at
pseudo-threshold due to $e.m.$ current conservation. And the new
MAMI measurement very well shows this fundamental symmetry.

\begin{figure}
\begin{center}
\includegraphics[width=6cm]{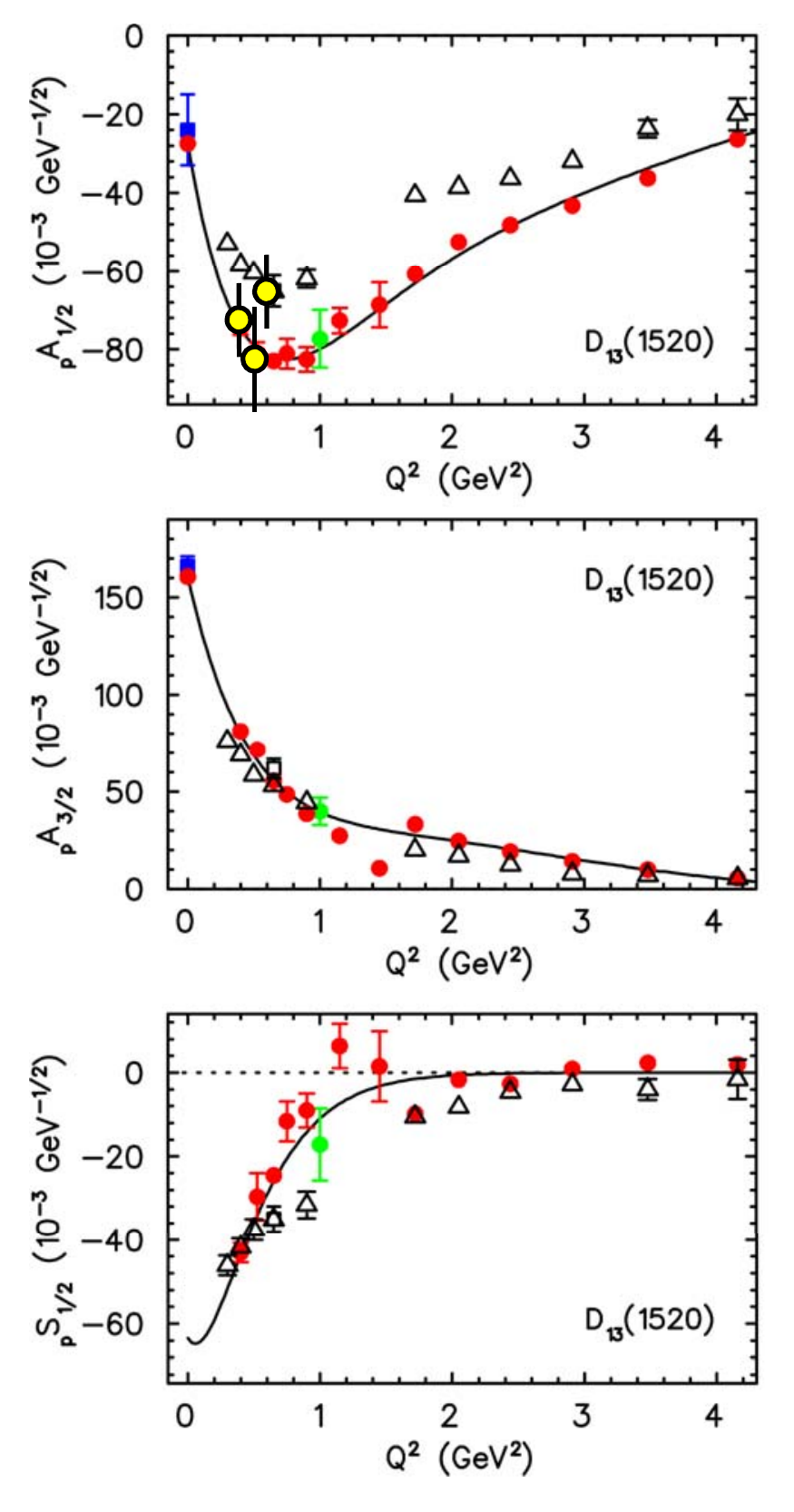}\hspace*{0mm}
\includegraphics[width=6cm]{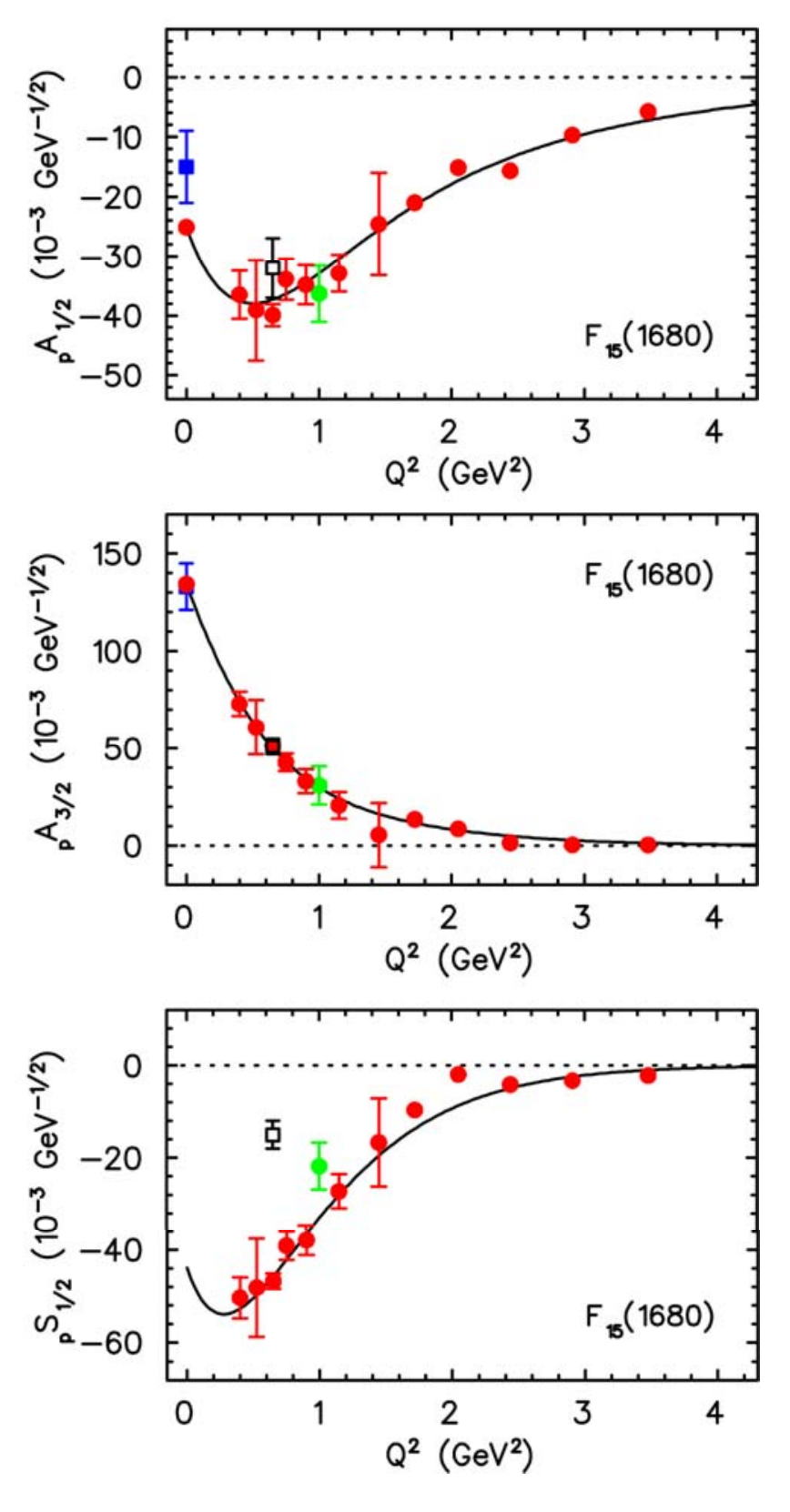}

\vspace*{3mm} \caption{\label{fig:d13_f15_ff}  Transverse
$_pA_{1/2},\, _pA_{3/2}$ and longitudinal $_pS_{1/2}$ transition
form factors of the D13 $N(1520)3/2^-$ and F15 $N(1680)5/2^+$
nucleon resonances. The curves show the MAID2008 parametrization.
The red circles are single-$Q^2$ results from our MAID analyses and
the black triangles from the JLab analysis~\cite{Aznauryan:2009mx}.
The yellow circles for $_pA_{1/2}(D13)$ show part of the 2012 CLAS
$\pi^+\pi-$ analysis~\cite{Mokeev:2012vsa}. For further details see
Ref.~\cite{Tiator:2011pw}. }
\end{center}
\end{figure}

In Fig.~\ref{fig:d13_f15_ff} the transverse $_pA_{1/2}(Q^2)$,
$_pA_{3/2}(Q^2)$ and the longitudinal $_pS_{1/2}(Q^2)$ form factors
for the transitions from the proton to the $N(1520)3/2^-$ (D13) and
$N(1680)5/2^+$ (F15) resonances are shown. Besides the
$\Delta(1232)3/2^+$ resonance transition, these two $N^*$
transitions are the most pronounced resonance structures in the
electroexcitation of the proton. Again as before, the MAID
parameterizations are in good agreement with the partial wave
analyses of MAID and CLAS. The only remarkable deviation is found
for the $_pA_{1/2}$ form factor to the D13 resonance. There,
however, a more recent analysis from $\pi^+ \pi^-$
electroproduction~\cite{Mokeev:2012vsa} is partly in better
agreement with our MAID analysis.

For both of these proton resonances the helicity non-conserving
amplitude $A_{3/2}$ dominates for real photons but with increasing
values of $Q^2$ it drops much faster than the helicity conserving
amplitude $A_{1/2}$.

\section{EtaMAID update 2017}

\begin{figure}
\begin{center}
\includegraphics[width=9cm]{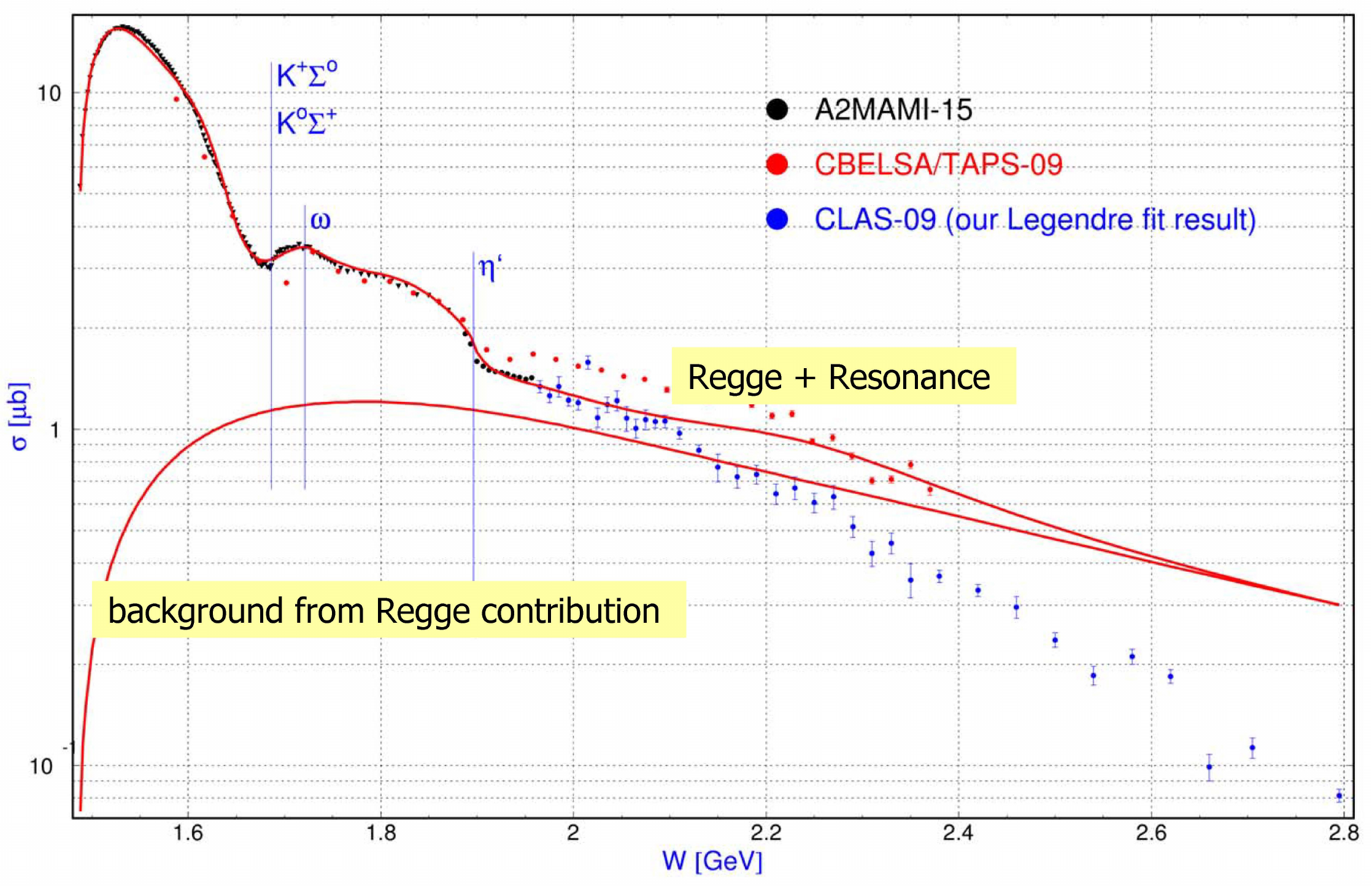}
%\hspace*{3mm}\includegraphics[width=6.5cm]{F15formfactors}
\vspace*{3mm} \caption{\label{fig:etamaid-1} Total cross section for
$\gamma p\to \eta p$ obtained with EtaMAID2017 in a
Regge-plus-Resonance approach. The separately shown Regge
contribution (lower red line) contains trajectory and cuts with
natural and un-natural parity.  The black data points are the recent
MAMI data~\cite{Kashevarov:2017kqb} and the red points show the Bonn
data of 2009~\cite{CBELSA_2009}. The blue data points, reaching to
the highest energies, are obtained from a Legendre fit of the
angular distributions of the 2009 CLAS data~\cite{CLAS_2009}.  }
\end{center}
\end{figure}

\begin{figure}
\begin{center}
\includegraphics[width=6cm]{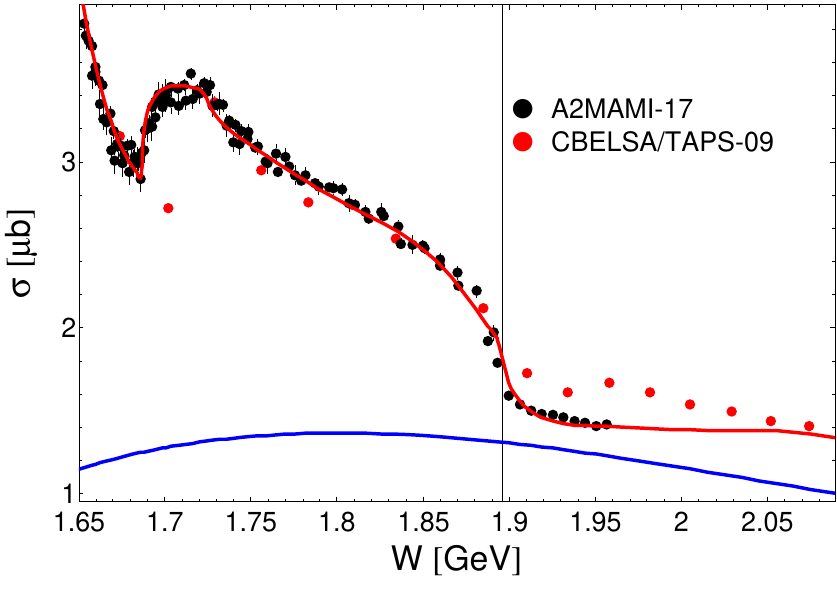}
\includegraphics[width=6cm]{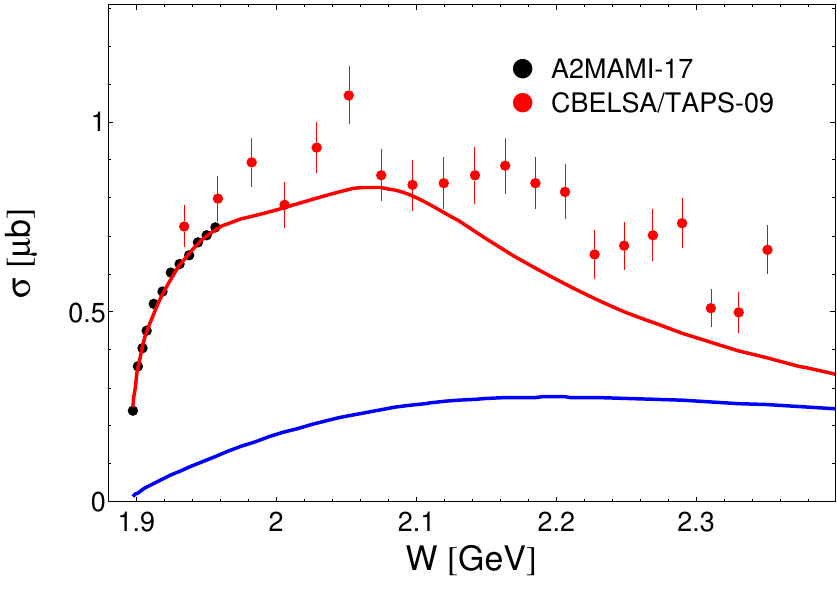}
\vspace*{3mm} \caption{\label{fig:etamaid-2} Total cross section for
$\gamma p\to \eta p$ (left panel) and $\gamma p\to \eta\prime p$
(right panel) obtained with EtaMAID2017 in a Regge-plus-Resonance
approach. The black data points are the recent MAMI
data~\cite{Kashevarov:2017kqb} and the red points show the Bonn data
of 2009~\cite{CBELSA_2009}. The separately shown Regge contribution
(lower blue lines) contain trajectory and cuts with natural and
un-natural parity.}
\end{center}
\end{figure}

With the advent of new polarization experiments in Mainz on eta
photoproduction~\cite{Akondi:2014ttg} and high precision
differential cross sections for eta and etaprime
photoproduction~\cite{Kashevarov:2017kqb}, the EtaMAID model is
being updated since 2015~\cite{Kashevarov:2016owq} but not yet
online. The current version EtaMAID2017 is shown in
Fig.~\ref{fig:etamaid-1} and Fig.~\ref{fig:etamaid-2} compared to
the MAMI total cross section data~\cite{Kashevarov:2017kqb}. In a
Regge-plus-Resonance (RPR) approach, the contributions of $N^*$
resonances are added to a Regge background, which dominates the
high-energy tail beyond the resonance region and becomes small near
threshold.

The new EtaMAID2017 model includes a non-resonant background, which
consists of the vector ($\rho$ and $\omega$) and axial-vector
($b_1$) exchanges in the $t$ channel, and $s$-channel $N^*$
excitations. Regge trajectories for the meson exchange in the $t$
channel were used to provide the correct asymptotic behavior at high
energies. In addition to the Regge trajectories, Regge cuts with
natural and unnatural parity were also
included~\cite{Kashevarov:2017vyl}.

The major role for the description of $\eta$ and $\eta'$
photoproduction is played by three $s$-wave resonances:
$N(1535)1/2^-$, $N(1650)1/2^-$, and $N(1895)1/2^-$, the latter of
which plays the key role in the features observed at the $\eta'$
threshold. Both the exact shape of the cusp in the $\eta$
photoproduction and the steepness of $\eta'$ photoproduction at
threshold are strongly correlated with the properties of
$N(1895)1/2^-$, allowing their extraction with good accuracy. Mainly
due to these data, a two-star status in PDG~\cite{PDG2016} has now
been raised to 4-star, the status for well established nucleon
resonances.

\section{Applications}

The MAID programs have been used in many different ways.\\

\noindent The first group are applications, where the theoretical
results can be obtained directly from the MAID web pages:
\begin{itemize}
\item comparison of cross sections and polarization observables with
experiment
\item predictions for new measurements and for experimental
proposals
\item comparison with different theoretical models and partial wave analysis
(PWA)
\item investigations of CGLN, helicity and invariant amplitudes
\item PWA with electric, magnetic and
charge multipoles
\item nucleon resonances $N^*$ and $\Delta^*$ in Breit-Wigner
parametrization
\item transverse and longitudinal transition form factors
\end{itemize}

\noindent The second group are indirect applications, where
numerical results of the MAID programs are used for further
calculations:
\begin{itemize}
\item T-matrix pole positions and residues of nucleon
resonances~\cite{Svarc:2014sqa,Tiator:2016btt}
\item transverse transition densities on the
light-front~\cite{Tiator:2008kd}
\item complete experiment analyses~\cite{Tiator:2017cde}
\item fixed-$t$ dispersion relations for meson photo- and
electroproduction~\cite{Pasquini:2007all}
\item dispersion relations in real and virtual Compton
scattering~\cite{Drechsel:2002ar}
\item Gerasimov-Drell-Hearn sum rule and spin structure
of the nucleon~\cite{Drechsel:2004ki}
\item nucleon polarizabilities and light-front
interpretation~\cite{Gorchtein:2009qq,Gasser:2015dwa}
\item two-photon exchange corrections to elastic $e$-$p$
scattering~\cite{Tomalak:2017shs}
\end{itemize}

\section{Summary and Outlook}

The MAID project is a collection of online programs that perform
real-time calculations for pseudoscalar meson photo- and
electroproduction. It started in 1998 with pion photo- and
electroproduction, and kaon, eta, etaprime and 2-pion production
followed soon. The most recent part is a chiral effective field
theory approach, where all free low-energy constants are fitted to
the world data of pion production in the threshold region. MAID has
been used in many different ways, for data analysis, experimental
proposals, for sum rules, dispersion theoretical calculations, in
particular for real, virtual and double-virtual Compton processes. A
lot of PhD students profitted very much from the easily available
MAID programs, and this gave us a lot of positive feedback.

Besides cross sections, the programs also provide all possible
polarization observables, including beam, target and recoil
polarization. For detailed investigations and partial wave analyses,
the programs also provide full sets of CGLN, helicity and invariant
amplitudes and electromagnetic multipoles, the partial waves of
photoproduction.

MAID has been updated only a few times, nevertheless it often has
proven predictive power after new experimental data became
available. However, with high statistics of recent data, especially
with MAMI cross section
data~\cite{Adlarson:2015byy,Kashevarov:2017kqb}, and with a lot of
new double-polarization observables from Mainz and Bonn, the
limitations of the MAID models became obvious, as can be seen in a
recent comparison of PWA from different data analysis
groups~\cite{Beck:2016hcy}.

In 2015 an EtaMAID update for $\gamma,\eta$ and $\gamma,\eta\prime$
was started with a much improved high-energy region, described by
Regge approaches~\cite{Kashevarov:2017vyl} and a resonance region
that now contains up to about 20 $N^*$
resonances~\cite{Kashevarov:2016owq,Kashevarov:2017kqb}. This part
will soon become online available.

The next updates are planned for kaon photoproduction, where a lot
of new and high-quality data became available during the last
decade. Furthermore, we also plan to update the pion photo- and
electroproduction MAID with new data and a series of new $N^*$ and
$\Delta^*$ resonances. So far, MAID2007 was limited to only 13
resonances, the full set of 4-star resonances below 2 GeV in 2007.

The pion electroproduction process has the biggest impact in
theoretical applications and will certainly be needed for a long
time. Whether the MAID service can be much longer provided is
currently unclear. Probably for another few years, but afterwards it
may become frozen for the future.

\begin{acknowledgements}
From the beginning, the MAID project was heavily supported by Prof.
Dieter Drechsel in Mainz, who also gave a lot of encouragement for
the development and for applications. A very big part of the
programs was developed by Sabit Kamalov from Dubna. In addition we
are also grateful to the contributions of Olaf Hanstein, Marc
Vanderhaeghen, Victor Kashevarov, Stefan Scherer and Marius Hilt
from Mainz, Shin Nan Yang and Wen Tai Chiang from Taipei, Terry Mart
from Depok (Indonesia), Cornelius Bennhold from George Washington
University and Alexander Fix from Tomsk. The MAID project was
supported by the Deutsche Forschungsgemeinschaft (SFB 201 and 1044).
\end{acknowledgements}

\end{document}